\documentclass[sigplan, screen, nonacm]{acmart}

\usepackage{microtype}

\usepackage{xparse}
\usepackage{xspace}

\usepackage{subfiles}

\usepackage{fancyvrb}
\usepackage{listings}

\usepackage{notationFonts}
\usepackage{separationLogic}
\usepackage{hoareTriples}
\usepackage{genMath}
\usepackage{notes}
\usepackage{substitution}

\AtBeginDocument{%
	}

\hypersetup{
	breaklinks=true,   
	colorlinks=true,   
	pdfusetitle=true,  
}

\begin{document}
	
	\title{Completeness Thresholds for Memory Safety of Array Traversing Programs}
	
	\author{Tobias Reinhard}
	\email{tobias.reinhard@kuleuven.be}
	\orcid{0000-0003-1048-8735}
	\affiliation{%
		\institution{imec-DistriNet, KU Leuven}
		\streetaddress{Celestijnenlaan 200A - bus 2402}
		\city{Leuven}
		\country{Belgium}
		\postcode{3001}
	}

	\author{Justus Fasse}
	\email{justus.fasse@kuleuven.be}
	\orcid{0009-0006-7383-7866}
	\affiliation{%
		\institution{imec-DistriNet, KU Leuven}
		\streetaddress{Celestijnenlaan 200A - bus 2402}
		\city{Leuven}
		\country{Belgium}
		\postcode{3001}
	}

	\author{Bart Jacobs}
	\email{bart.jacobs@kuleuven.be}
	\orcid{0000-0002-3605-249X}
	\affiliation{%
		\institution{imec-DistriNet, KU Leuven}
		\streetaddress{Celestijnenlaan 200A - bus 2402}
		\city{Leuven}
		\country{Belgium}
		\postcode{3001}
	}



	\begin{abstract}

We report on intermediate results of -- to the best of our knowledge -- the first study of \emph{completeness thresholds} for (partially) bounded memory safety proofs.
Specifically, we consider heap-manipulating programs that iterate over arrays without allocating or freeing memory.
In this setting, we present the first notion of completeness thresholds for program verification which reduce \emph{unbounded} memory safety proofs to (partially) \emph{bounded} ones.
Moreover, we demonstrate that we can characterise completeness thresholds for simple classes of array traversing programs.
Finally, we suggest avenues of research to scale this technique theoretically, i.e., to larger classes of programs (heap manipulation, tree-like data structures), and practically by highlighting automation opportunities.

	\end{abstract}

	
	
	
	\maketitle

	\DefineVerbatimEnvironment{code}{BVerbatim}{baseline=t}

	\NewDocumentCommand{\sizeVar}{}{\metaVar{s}}
	\NewDocumentCommand{\sizeLow}{}
		{\metaVar{\sizeVar_-}}
		\NewDocumentCommand{\sizeHigh}{}
		{\metaVar{\sizeVar_+}}
	\NewDocumentCommand{\cmdVar}{}{\metaVar{c}}
	\NewDocumentCommand{\arrayVar}{}{\metaVar{a}}
	\NewDocumentCommand{\indexVar}{}{\metaVar{i}}
	\NewDocumentCommand{\leftRangeBound}{}{\metaVar{L}}
	\NewDocumentCommand{\rightRangeBound}{}{\metaVar{R}}
	\NewDocumentCommand{\sizeBoundVar}{}{\metaVar{S}}
	\NewDocumentCommand{\depthBoundVar}{}{\metaVar{D}}
	\NewDocumentCommand{\zVar}{}{\metaVar{Z}}

	\NewDocumentCommand{\travPattern}{o o o}
		{\ensuremath{
			\mathit{trav}
			_{
				\IfValueTF{#1}{#1}{\leftRangeBound},
				\IfValueTF{#2}{#2}{\rightRangeBound}
			}
			^{
				\IfValueTF{#3}{#3}{\zVar}
			}
		}\xspace}
	\NewDocumentCommand{\travPatternVars}{o o o m}
		{\metaVar{\travPattern[#1][#2][#3](#4)}}
	
	\NewDocumentCommand{\travPatternClass}{}
		{\ensuremath{
			\mathit{Trav}
		}\xspace}

	\NewDocumentCommand{\compProg}{o o o}
		{\ensuremath{
			\mathit{comp}
			_{
				\IfValueTF{#1}{#1}{\leftRangeBound},
				\IfValueTF{#2}{#2}{\rightRangeBound}
			}
			^{
				\IfValueTF{#3}{#3}{\zVar}
			}
		}\xspace}
	\NewDocumentCommand{\compProgVars}{o o o m}
		{\metaVar{\compProg[#1][#2][#3](#4)}}
	
	\NewDocumentCommand{\compProgClass}{}
		{\ensuremath{
			\mathit{Comp}
		}\xspace}

	\NewDocumentCommand{\sumPattern}{o o o}
		{\ensuremath{
			\mathit{sum}
			_{
				\IfValueTF{#1}{#1}{\leftRangeBound},
				\IfValueTF{#2}{#2}{\rightRangeBound}
			}
			^{
				\IfValueTF{#3}{#3}{\zVar}
			}
		}\xspace}
	\NewDocumentCommand{\sumPatternVars}{o o o m}
		{\metaVar{\sumPattern[#1][#2][#3](#4)}}

	\NewDocumentCommand{\ctVar}{}{\metaVar{Q}}
	\NewDocumentCommand{\ctElem}{o}
		{\ensuremath{
			\metaVar{q}
			\IfValueT{#1}{
				_{#1}
			}
		}\xspace}
	\NewDocumentCommand{\ctElemLow}{}{\ctElem[-]}
	\NewDocumentCommand{\ctElemHigh}{}{\ctElem[+]}
	\NewDocumentCommand{\domVar}{}{\metaVar{X}}
	\NewDocumentCommand{\constraintSetVar}{}{\metaVar{K}}

	\NewDocumentCommand{\memsafe}{m}
		{\ensuremath{
			\texttt{memsafe}(#1)
		}\xspace}

	\NewDocumentCommand{\vcVar}{}{\metaVar{vc}}

	\NewDocumentCommand{\vcFalse}{}
		{\ensuremath{
			\vcVar_{\text{false}}
		}\xspace}

	\NewDocumentCommand{\vcTrav}{}
		{\ensuremath{
			\vcVar_{\text{trav}}
		}\xspace}

	\NewDocumentCommand{\vcSum}{o}
		{\ensuremath{
			\vcVar_{\text{sum}}
				\IfValueT{#1}{
					^{#1}
				}
		}\xspace}

	\NewDocumentCommand{\preVar}{}{\metaVar{A}}
	\NewDocumentCommand{\postVar}{}{\metaVar{B}}
	\NewDocumentCommand{\memAssVar}{}{\metaVar{M}}

	\NewDocumentCommand{\arrayPredName}{}
		{\fixedPredNameFont{array}\xspace}
	\NewDocumentCommand{\arrayPred}{O{\arrayVar} O{\sizeVar}}
		{\ensuremath{
			\arrayPredName(#1, #2)
		}\xspace}
	
	\NewDocumentCommand{\unrelStructVar}{}{\metaVar{y}}
	\NewDocumentCommand{\unrelSizeVar}{}
		{\metaVar{k}}
	\NewDocumentCommand{\dataPredName}{}
		{\fixedPredNameFont{complex\_data}}
	\NewDocumentCommand{\unrelDataPred}{O{\unrelStructVar} O{\unrelSizeVar}}
		{\ensuremath{
			\dataPredName(#1, #2)
		}\xspace}
	\NewDocumentCommand{\vcComp}{}{\metaVar{\vcVar_{comp}}}

	\NewDocumentCommand{\resVar}{}{\metaVar{r}}
	\NewDocumentCommand{\wlpName}{}{\fixedFuncNameFont{wlp}\xspace}
	\NewDocumentCommand{\wlp}{m O{\resVar} m}
		{\ensuremath{
			\wlpName
			(#1, \lambda #2.\ #3)
		}\xspace}

	\NewDocumentCommand{\nVar}{}{\metaVar{n}}

	\NewDocumentCommand{\vcLine}{m}
		{\ensuremath{
			(\texttt{\##1})
		}\xspace}

	\NewDocumentCommand{\vcTransComment}{m}
		{\texttt{// #1}}

	\NewDocumentCommand{\subAssVar}{o}
		{\ensuremath{
			\metaVar{A}
			\IfValueT{#1}{
				_{#1}
			}
		}\xspace}
	
	\NewDocumentCommand{\freeVars}{m}
		{\ensuremath{
			\fixedFuncNameFont{freeVars}(#1)
		}\xspace}

	\NewDocumentCommand{\deleteBlock}{m}{}

	\definecolor{codegreen}{rgb}{0,0.6,0}
\definecolor{codegray}{rgb}{0.5,0.5,0.5}
\definecolor{codepurple}{rgb}{0.58,0,0.82}
\definecolor{backcolour}{rgb}{0.95,0.95,0.92}

\lstdefinestyle{WHILEstyle}{
    backgroundcolor=\color{backcolour},   
    commentstyle=\color{codegreen},
    keywordstyle=\color{magenta},
    numberstyle=\tiny\color{codegray},
    stringstyle=\color{codepurple},
    basicstyle=\ttfamily\footnotesize,
    breakatwhitespace=false,         
    breaklines=true,                 
    captionpos=b,                    
    keepspaces=true,                 
    numbers=left,                    
    numbersep=5pt,                  
    showspaces=false,                
    showstringspaces=false,
    showtabs=false,                  
    tabsize=2
}
\lstset{style=WHILEstyle}

	\section{Introduction}\label{sec:Intro}

\paragraph{Unbounded vs Bounded Proofs}
Many techniques have been developed to convince ourselves of the trustworthiness of software.
A fundamental pillar for any higher-level property is memory safety.
In memory-unsafe languages the burden of proof lies with the programmer.
Yet, it remains hard to prove and in general requires us to write tedious, inductive proofs.
One way to automate the verification process is to settle on bounded proofs and accept bounded guarantees.

Consider a program \cmdVar that searches through an array of size \sizeVar.
An unbounded memory safety proof for \cmdVar would yield that the program is safe for any possible input, in particular for any array size, i.e., 
$\forall \sizeVar.\ \texttt{memsafe}(\cmdVar(\sizeVar))$.
A bounded proof that only considers input sizes up to ten would only guarantee that the program is safe for any such bounded array, i.e., 
$\forall \sizeVar \leq 10.\ \texttt{memsafe}(\cmdVar(\sizeVar))$.

\paragraph{Completeness Thresholds}
Approximating unbounded\\ proofs by bounded ones is a technique often used in model checking.
Hence, the relationship between bounded and unbounded proofs about finite state transition systems has been studied extensively~\cite{Biere1999SymbolicMC, Clarke2004CompletenessAC, Kroening2003EfficientCO, Bundala2012OnTM, Abdulaziz2018FormallyVA, Heljanko2005IncrementalAC, Awedh2004ProvingMP, McMillan2003InterpolationAS}.
For a finite transition system $T$ and a property of interest $\phi$, a \emph{completeness threshold} is any number $k$ such that we can prove $\phi$ by only examining path prefixes of length $k$ in $T$, i.e., $T \models_k \phi \Rightarrow T \models \phi$~\cite{Clarke2004CompletenessAC}~\footnote{
	Note that the term completeness threshold is used inconsistently in literature.
	Some papers such as~\cite{Clarke2004CompletenessAC} use the definition above, according to which completeness thresholds are not unique.
	Others like~\cite{Kroening2003EfficientCO} define them as the minimal number $k$ such that $T \models_k \phi \Rightarrow T \models \phi$, which makes them unique.
}.
Over the years, various works characterised over-approximations of least completeness thresholds for different types of properties~$\phi$.
These over-approximations are typically described in terms of key attributes of the transition system $T$, such as the \emph{recurrence diameter} (longest loop-free path)~\cite{Kroening2003EfficientCO}.
For instance, consider the class of global safety properties of the form $Gp$ for finite transition systems $T$, where $p$ is a local property.
We know that the smallest completeness threshold for this class expressible solely in terms of $T$'s diameter is exactly $\texttt{diam}(T)$~\cite{Biere1999SymbolicMC, DBLP:conf/cav/KroeningOSWW11}.
As safety property of the form $Gp$, this also applies to memory safety of finite transition systems.

In general, heap-manipulating programs' state space can be infinite.
That is because the program's input data can be arbitrarily large and because executions can be arbitrarily long.
Therefore, the key attributes described above will generally be infinite as well.
This vast structural difference between the programs we are interested in and the transition systems for which completeness thresholds have been studied prevents us from reusing any of the existing definitions or results.

	\section{Limitations of Bounded Proofs}\label{sec:LimationsOfBoundedProofs}

\paragraph{Bounded Model Checking}
Generally, if we want un\-boun\-ded memory safety guarantees, we have to consider all possible input sizes and all possible executions.
This is often hard and requires us to write tedious inductive proofs.
An alternative is to give up on the idea of unbounded guarantees and to settle for bounded ones.
One approach that has proven useful during development of critical software is bounded model checking~(BMC)~\cite{Chong2020CodeLeveMC}.

The underlying idea is to approximate the original verification problem by a finite model that we can check automatically.
With this approach, we choose a size bound \sizeBoundVar and only consider inputs with sizes up to \sizeBoundVar.
Further, we also only check finite execution prefixes.
A common approach is to unwind loops and recursion up to a certain depth.

The intuition behind this approach is that if the program contains errors, they likely already occur for small input sizes and early loop iterations.
As long as BMC does not perform abstraction~\cite{Clarke2000CounterExGuidedRefinement}, all reported counterexamples are real bugs.
However, we should be careful not to forget that this way we only obtain a bounded proof yielding bounded~guarantees.

\begin{figure}
	\raggedright
	\begin{tabular}{l l l}
		$\left.
		\begin{minipage}{4cm}
\begin{lstlisting}
for i in [L : s-R] do
	!a[i+Z]
\end{lstlisting}
		\end{minipage}
		\quad\right\}
		$
		=: 
		&\travPatternVars{a, s}
	\end{tabular}
	\\
	\begin{tabular}{l l l}
		\travPatternClass
		&:=
		&$\{ \travPattern \quad | \quad \leftRangeBound, \rightRangeBound, \zVar \in \Z \}$
	\end{tabular}
	\caption{
		Class \travPatternClass of programs \travPattern traversing an array \arrayVar of size \sizeVar, attempting to read elements.
		$\leftRangeBound, \rightRangeBound, \zVar$ \mbox{are constants}.
	}
	\label{fig:ProgClass}
\end{figure}

\paragraph{Array Traversal Pattern}
Consider the class of programs \travPatternClass presented in Fig.~\ref{fig:ProgClass} in a WHILE language with pointer arithmetic.
Given a pointer \arrayVar and a variable \sizeVar, such that \arrayVar points to an array of size \sizeVar,
each program iterates through the array and attempts to read elements.
The class models a basic programming pattern and common off-by-n errors~\cite{cwe193}. 
We use upper case letters for constants and lower case letters for (program) variables.
A program \travPattern from this class iterates from $\indexVar = \leftRangeBound$ to $\indexVar = \sizeVar - \rightRangeBound$ (bounds incl.) and attempts to read the array at index $\indexVar + \zVar$.
We use $!x$ to express accesses to a heap location $x$.
Whether memory errors occur for a concrete instance \travPattern depends on how the constants \leftRangeBound, \rightRangeBound, \zVar are chosen.
We use it as minimal example \mbox{throughout this paper}.

\paragraph{What Could Go Wrong with Bounded Proofs?\!\!}\
To illustrate the issue, let us use BMC to check various instances of the array traversal pattern:
(i)~traversal of the entire array: \travPattern[0][1][0],
(ii)~traversal of the array with accesses offset by two from the index: \travPattern[0][1][2] and
(iii)~an additional reduction of the index variable's upper bound by one: \travPattern[0][2][2].
It is easy to see that (i) is memory-safe while (ii) and (iii) are not.
However, before we run a model checking algorithm we have to choose appropriate bounds.
The pattern we are looking at is quite simple.
So, we choose size bound $\sizeBoundVar = 1$ and unwinding depth $\depthBoundVar = 1$ for the BMC procedure.
Note that the latter effectively means:
we do not restrict the loop depth for the input sizes we chose.

For the standard variant (i)~we cannot find any errors within the bounds.
This is fine because the program is safe.
In variant (ii)~array accesses $\arrayVar[\indexVar+2]$ are incorrectly shifted to the right.
This already leads to an out-of-bounds error for arrays of size 1.
This size falls within our chosen bounds, so BMC reports this error and we can correct it.
Finally, (iii)'s reduction of the index variable's range to $[0,\sizeVar - 2]$ means that the program only performs loop iterations for arrays of size $\sizeVar \geq 2$.
Consequently, it is trivially safe for the sizes $0$ and $1$.
These are the sizes our bounded proof explores.
Hence, BMC does not report any errors and leads us to wrongly believe that \travPattern[0][2][2] is safe.

	\section{Completeness Thresholds}\label{sec:CTs}

As illustrated above, bounded proofs are in general unsound approximations of unbounded proofs.
A concrete approximation is sound iff we choose the bounds large enough, such that we can be sure that we do not miss any errors.
We focus on bounding input sizes (in our examples array sizes), ignoring loop bounds that do not depend on these~parameters.

Recall from \S~\ref{sec:Intro} that completeness thresholds are a concept from model checking of finite transition systems~\cite{Clarke2004CompletenessAC}.
We borrow this terminology and apply it to memory safety verification.
Hence, for a program $\cmdVar(x)$ with input parameter $x \in \domVar$, we call any subdomain $\ctVar \subseteq \domVar$ a \emph{completeness threshold}~(\emph{CT}) for $x$ in $\cmdVar$ if we can prove memory safety of \cmdVar by only considering inputs from \ctVar, i.e., 
$
\forall x \in \ctVar.\ 
  \texttt{memsafe}(\cmdVar(x))
$
$\Rightarrow
\forall x \in \domVar.\ 
  \texttt{memsafe}(\cmdVar(x))
$.

\paragraph{Intuitive CT Extraction}
Returning to our example class of programs \travPatternClass implementing the array traversal pattern.
Some of these are memory safe, some are not.
So, let us try to compute completeness thresholds for these programs.
First, let's take a look at the errors that might occur.
This gives us an idea which sizes a sound bounded proof must cover.
Any instance \travPatternVars{\arrayVar, \sizeVar} iterates ascendingly $i = \leftRangeBound, \dots, \sizeVar-\rightRangeBound$ and accesses $\arrayVar[i+\zVar]$.
For sizes \sizeVar that cause the ascending range $\leftRangeBound, \dots, \sizeVar-\rightRangeBound$ to be empty, we do not execute the loop at all.
Any such run is trivially memory safe.
Therefore, any meaningful bounded proof of \travPatternVars{\arrayVar, \sizeVar} must include sizes~\sizeVar with 
$\{\leftRangeBound, \dots, \sizeVar-\rightRangeBound\} \neq \emptyset$, i.e.,
$\sizeVar \geq \leftRangeBound+\rightRangeBound$.

Suppose $\sizeVar \geq \leftRangeBound+\rightRangeBound$.
An error occurs if the index $\indexVar+\zVar$ violates the array bounds, i.e., if $\indexVar+\zVar < 0$ or $\indexVar+\zVar \geq \sizeVar$.
Taking the index range into account, we see that we get an error if
$\leftRangeBound+\zVar < 0$ 
or 
$\sizeVar-\rightRangeBound+\zVar \geq \sizeVar$ holds.
We can simplify the latter to $\zVar-\rightRangeBound \geq 0$.

Note that neither $\leftRangeBound+\zVar < 0$ nor $\zVar-\rightRangeBound \geq 0$ depend on the array size \sizeVar.
This means that as long as we focus on sizes above the threshold $\sizeVar \geq \leftRangeBound+\rightRangeBound$,
the concrete choice of \sizeVar does not influence whether an error occurs or not.
In other words, it suffices for our bounded proof to only check a single (arbitrarily chosen) size
$\ctElem \geq \leftRangeBound+\rightRangeBound$
and then we can extrapolate the result, i.e.,
$$
  \forall \arrayVar.\
	  \memsafe{\travPatternVars{\arrayVar, q}}
  \ \Rightarrow\
  \forall \sizeVar.
  \forall \arrayVar.\
    \memsafe{\travPatternVars{\arrayVar, \sizeVar}}
$$

Hence, any set $\{q\}$ for $q \geq \leftRangeBound+\rightRangeBound$ is a CT for the array size parameter \sizeVar in \travPattern.
We just found a uniform characterization of CTs for the entire class~\travPatternClass.
Note that $\{q\}$ is not necessarily the smallest CT.
For safe instances such as \travPattern[0][1][0], the empty set $\emptyset$ is a valid CT as well.

\paragraph{Our Approach}
We study CTs for $x$ in $\cmdVar(x)$ by studying its \emph{verification condition} (VC) . 
The latter is an automatically generated logical formula of the form
$\forall x \in \domVar.\ \vcVar(x)$
and proving it entails memory safety of $\cmdVar(x)$ for all choices of $x$.
Next, we currently simplify $\vcVar(x)$ by hand until it becomes clear how the choice of $x$ affects the validity of $\vcVar(x)$.
Knowing this allows us to partition the domain into 
$\domVar = \bigcup \ctVar_i$.
For each subdomain we get $\vcVar_i(x) = \forall x \in \ctVar_i.\ \vcVar(x)$.
If possible, we simplify each $\vcVar_i'(x)$ into $\vcVar_i''(x)$ based on the restricted subdomain $\ctVar_i$ with the goal to eliminate occurrences of $x$.
If $\vcVar_i''$ does not mention $x$ we pick any element of $\ctVar_i$ as representative $\ctVar_i'$.
Otherwise, $\ctVar_i' = \ctVar_i$.
Hence, $\bigcup \ctVar_i'$ is a CT for $x$ in $\cmdVar(x)$.
In the following we elaborate this in more detail.

\subsection{Approximating CTs via Verification Conditions}
\label{sec:CTs:ApproxCTsViaVCs}

Now that we have an intuition for the CTs of \travPatternClass, let's turn our informal argument from above into a formal one.
Formal definitions of the language and logic we consider and proofs for the presented lemmas can be found in the technical report~\cite{Reinhard2023ct4ms-TR}.

\paragraph{Hoare Triples}
We use Hoare triples~\cite{Hoare1968HoareLogic} to express program specifications.
A triple \hoareTriple{\preVar}{\cmdVar}{\postVar} expresses that the following properties hold for every execution that starts in a state which satisfies precondition~\preVar:
Firstly, the execution does not encounter any runtime errors.
Secondly, it either
(i)~does not terminate or
(ii)~it terminates in a state complying with postcondition~\postVar.

In this work, we study the memory safety of programs that do not change the shape of the data structures they process.
Hence, we choose preconditions that merely describe the memory layout of the data structures which our programs receive as input.
For the array traversal program, we choose the predicate \arrayPred as precondition, which expresses that \arrayVar points to a contiguous memory chunk of size \sizeVar.
Given that our target programs do not change the memory layout, specifications simplify to \hoareTriple{\preVar}{\cmdVar}{\preVar}.
For the array traversal we get
\hoareTriple{\arrayPred}{\travPattern}{\arrayPred}.

\begin{definition}[Completeness Thresholds for Programs]
  Let \hoareTriple{\preVar}{\cmdVar}{\postVar} be a program specification containing a free variable $x$ with domain \domVar.
  We call a subdomain $\ctVar \subseteq \domVar$ a \emph{completeness threshold} for $x$ in \hoareTriple{\preVar}{\cmdVar}{\postVar} if
  $$
    \models 
      \forall x \in \ctVar.\
      \hoareTriple{\preVar}{\cmdVar}{\postVar}
    \quad\Rightarrow\quad
    \models
      \forall x \in \domVar.\
      \hoareTriple{\preVar}{\cmdVar}{\postVar}
  $$
\end{definition}
We omit spelling out the pre- and postconditions when they are clear from the context.
Instead we say that \ctVar is a completeness threshold for $x$ in program \cmdVar.

\paragraph{Separation Logic}
We use a first-order affine/intuitionistic separation logic with recursion predicates~\cite{DBLP:journals/cacm/OHearn19,OHearn2001LocalRA,Reynolds2002SeparationLA}
to describe memory.
Since we focus on heap-manipulating programs, we use assertions to describe heaps.
Separation logic comes with a few special operators:
(i)~The points-to chunk \slPointsTo{x}{v} describes a heap containing a location $x$ which holds the value~$v$.
We write \slPointsToAny{x} to express that we do not care about the value stored in the heap cell.
(ii)~The separating conjunction $\assVar_1 \slStar \assVar_2$ expresses that $\assVar_1$ and $\assVar_2$ describe disjoint heaps.
Hence, $\slPointsToAny{x} \slStar \slPointsToAny{y}$ implies that $x \neq y$.
(iii)~The separating implication $\assVar_1 \slWand \assVar_2$ can be read as $\assVar_2$ \emph{without} $\assVar_1$.
That is, combining the heap described by $\assVar_1 \slWand \assVar_2$ with a disjoint heap described by $\assVar_1$ yields a heap compliant with $\assVar_2$.
(iii)~In our logic, the persistence modality~$\slPersistent \assVar$ means that~\assVar does not describe resources and hence holds under the empty heap (cf.\ \cite{Jung2018IrisGroundUp}).

We assume that \arrayPredName denotes a (recursively defined) predicate, such that for every fixed size \sizeVar, we can express it as iterated separating conjunction: 
$\displaystyle
  \arrayPred 
  \equiv
  \slBigStar_{0 \leq k < s} \slPointsToAny{\arrayVar[k]}
$.

\paragraph{Verification Conditions}
A common way to verify programs is via verification conditions~\cite{Flanagan2001AvoidingExpExplosionVC, Parthasarathy2021VCGenerator}.
For any specification \hoareTriple{\preVar}{\cmdVar}{\postVar}, a \emph{verification condition}~(\emph{VC}) is any logical formula \vcVar, such that we can verify \hoareTriple{\preVar}{\cmdVar}{\postVar} by proving \vcVar, i.e.,
$
\models \vcVar
\ \Rightarrow\, \
\models \hoareTriple{\preVar}{\cmdVar}{\postVar}
$.

\begin{definition}[Verification Condition]
  We call an assertion \assVar a \emph{verification condition} for \hoareTriple{\preVar}{\cmdVar}{\postVar} if 
  $$
    \models \assVar
    \quad\Rightarrow\quad
    \models \hoareTriple{\preVar}{\cmdVar}{\postVar}.
  $$
\end{definition}

\begin{definition}[Completeness Thresholds for Assertions]
  Let \assVar be an assertion with a free variable $x$ of domain $X$.
  We call a subdomain $\ctVar \subseteq X$ a \emph{completeness threshold} for $x$ in \assVar if
  $$
    \models \forall x \in \ctVar.\ \assVar
    \quad\Rightarrow\quad
    \models \forall x \in X.\ \assVar.
  $$
\end{definition}

Consider a specification 
\hoareTriple{\preVar}{\cmdVar}{\postVar}
with a free variable $x \in X$ and a corresponding VC 
$\forall x \in X.\ \vcVar$.
Suppose we get a completeness threshold \ctVar for $x$ in \vcVar.
Knowing this threshold reduces correctness of the specification to the bounded VC, i.e.,
$
\models \forall x \in \ctVar.\ \vcVar 
\Rightarrow 
\hoareTriple{\preVar}{\cmdVar}{\postVar}
$.
That is, we can derive unbounded guarantees from a bounded proof.
We usually omit quantification domains when they are clear from the~context.

\paragraph{Weakest Liberal Preconditions}
A common way to generate VCs is via weakest liberal preconditions~\cite{Flanagan2001AvoidingExpExplosionVC, Dijsktra1976DisciplineOfProgramming}.
For any program \cmdVar and postcondition \postVar, the \emph{weakest liberal precondition}
\wlp{\cmdVar}{\postVar}
is an assertion for~which
$$
  \forall \preVar.\ \
  \big(\
    \preVar \models \wlp{\cmdVar}{\postVar}
    \ \Rightarrow\ \
    \models \hoareTriple{\preVar}{\cmdVar}{\postVar}
  \ \big)
$$
holds.
That is, if the weakest liberal precondition holds for the starting state,
then \cmdVar does either not terminate or it terminates in a state complying with postcondition \postVar.
In particular, no memory error occurs during the execution.
The canonical VC for \hoareTriple{\preVar}{\cmdVar}{\postVar} is 
$\forall \overline{x}.\ \preVar \rightarrow \wlp{\cmdVar}{\postVar}$
where $\overline{x}$ is the tuple of variables occuring freely in \preVar, \cmdVar and \postVar.

\paragraph{Limitations of CTs}
In general, VCs are over-approxima\-tions.
Hence, CTs derived from a VC do not always apply to the corrresponding program.
Consider the specification
\hoareTriple
  {\arrayPred}
  {\travPattern[0][2][2]}
  {\arrayPred}
and any unsatisfiable assertion
$\vcFalse \equiv \slFalse$.
Then 
$\forall \sizeVar \in \N.\ \vcFalse$ 
is an over-ap\-prox\-i\-ma\-ting VC, since
$
  \slFalse 
  \Rightarrow\
  \models
    \hoareTriple
      {\arrayPred}
      {\travPattern[0][2][2]}
      {\arrayPred}
$.
As discussed in \S~\ref{sec:LimationsOfBoundedProofs}, a bounded proof that only covers sizes $0, 1$ does not discover the errors in \travPattern[0][2][2].
Hence, the set $\{0\}$ is not a CT for $x$ in
\hoareTriple
  {\arrayPred}
  {\travPattern[0][2][2]}
  {\arrayPred}.
However, $\{0\}$ is a CT for \sizeVar in \vcFalse, since
$
  \forall \sizeVar \in \{0\}.\ \vcFalse
  \equiv
  \slFalse
$
and 
$
  \slFalse 
  \Rightarrow \
  \models \forall \sizeVar \in \N.\ \vcFalse
$.
We see that we can only transfer a CT \ctVar for some variable $x$ derived from a VC \vcVar to the corresponding program, if \vcVar does not over-approximate with regard to $x$.
(We are currently studying this connection.)
Note that this is, however, the case for the examples we discuss in this paper.
Though, even if \vcVar does over-approximate, CT \ctVar still applies to any proof considering a property at least as strong as \vcVar.

\deleteBlock{
  \begin{definition}[Pecision]
    Let \assVar and \hoareTriple{\preVar}{\cmdVar}{\postVar} be an assertion and specification, respectively, with a free variable $x$ of domain \domVar.
    We call \assVar \emph{precise} in $x$ for \hoareTriple{\preVar}{\cmdVar}{\postVar} if the following holds for all values $v \in \domVar$:
    $$
      \models
        \subst{\assVar}{x}{v}
      \quad\Rightarrow\quad
      \models
        \subst
          {\big(\hoareTriple{\preVar}{\cmdVar}{\postVar}\big)}
          {x}
          {v}
    $$
  \end{definition}
}

As the name suggests, VCs derived from weakest preconditions yield very weak properties.
It is reasonable to assume that often a bounded proof would imply a bounded version of the \wlpName-based VC.
Therefore, it is reasonable to use them during our study of CTs.

\paragraph{Extracting CTs via VCs}
Applying the above approach to our specification
\hoareTriple
  {\arrayPred}
  {\travPatternVars{\arrayVar, \sizeVar}}
  {\arrayPred},
we get a VC
$\forall \arrayVar.\ \forall \sizeVar.\ \vcTrav$,
where $\vcTrav$ is as follows~\footnote{
  The weakest precondition calculus requires us to annotate loops with loop invariants.
  In the setting we study, the initial memory layout is invariant under the program's execution.
  The preconditions we consider describe exactly the initial memory layout, nothing else.
  Hence, we can reuse preconditions as loop invariants during the \wlpName computation.
}:
$$
\begin{array}{p{0.55cm} p{0.05cm} l l}
  \!\vcTrav  
    &:=
    &\arrayPred \rightarrow
    &\vcLine{1}
    \\
    &&\quad
      \arrayPred
    &\vcLine{2}
    \\
    &&\quad
      \slStar\ \slPersistent(
        \forall \indexVar.\
          (\leftRangeBound \leq \indexVar \leq \sizeVar - \rightRangeBound)
          \wedge
          \arrayPred
          \slWand
    &\vcLine{3}
    \\
    &&\quad
      \quad\quad\quad\quad
          \exists v.\
          \slPointsTo{\arrayVar[i+\zVar]}{v}
          \wedge
          \arrayPred
      )
      \!\!
    \\
    &&\quad
      \slStar\ (\arrayPred \slWand \arrayPred)
    &\vcLine{4}
\end{array}
$$
Here, 
\vcLine{1}~is the precondition describing our memory layout.
The separating conjunction \vcLine{2}-\vcLine{4} is the weakest precondition derived from our specification and \vcTrav says that it should follow from precondition~\vcLine{1}.

The weakest precondition states that the memory layout must stay invariant under the loop execution.
\vcLine{2}~says that it should hold before the loop starts.
\vcLine{3}~demands that every loop iteration preserves the layout, i.e., that the layout description is a loop invariant.
\vcLine{4}~states this invariant implies the postcondition from our specification, which, again, is the unchanged memory layout.

Remember that we use an affine separation logic.
Clearly, this VC contains many trivially obsolete parts.
We can simplify \vcTrav to $\vcVar_1$:

$$
\begin{array}{l l l l}
  \vcTrav
  &\equiv
  &\vcTransComment{Eliminate \vcLine{2}, \vcLine{4}}
  \\
  &&\arrayPred \rightarrow
  \\
  &&\quad
    \slPersistent(
      \forall \indexVar.\
        (\leftRangeBound \leq \indexVar \leq \sizeVar - \rightRangeBound)
        \wedge
        \arrayPred
        \slWand
  \\
  &&\quad
    \quad\quad\quad\quad
        \exists v.\
        \slPointsTo{\arrayVar[i+\zVar]}{v}
        \wedge
        \arrayPred
    )
  \\
  &\equiv
  &\vcTransComment{Persistency makes pre.\! obsolete}
  \\
  &&\quad
    \slPersistent(
      \forall \indexVar.\
        (\leftRangeBound \leq \indexVar \leq \sizeVar - \rightRangeBound)
        \wedge
        \arrayPred
        \slWand
  \\
  &&\quad
    \quad\quad\quad\quad
        \exists v.\
        \slPointsTo{\arrayVar[i+\zVar]}{v}
        \wedge
        \arrayPred
    )
    \\
    &\equiv
    &\vcTransComment{
      \arrayPred equiv.\! to
      $\displaystyle\slBigStar_{0\leq k<s} \slPointsToAny{\arrayVar[k]}$
    }
    \\
    &&\quad
        \forall \indexVar.\
          (\leftRangeBound \leq \indexVar \leq \sizeVar - \rightRangeBound)
          \rightarrow
          (0 \leq \indexVar + \zVar < \sizeVar)
  \\
  &=:
  &\vcVar_1
\end{array}
$$

This equivalent VC $\forall \arrayVar.\ \forall \sizeVar.\ \vcVar_1$ does reflect the intuition we developed when analysing the program informally:
For sizes $\sizeVar < \leftRangeBound + \rightRangeBound$, the program does not perform any loop iterations and hence it is trivially memory safe.
For bigger arrays, a memory error occurs iff index $\indexVar+\zVar$ violates the array bounds.

We can justify this intuition by partitioning the domain of \sizeVar into 
$
  \N = 
  \{0, \dots, \leftRangeBound + (\rightRangeBound - 1)\} 
  \cup 
  \{\leftRangeBound + \rightRangeBound, \dots\}
$.
Let's analyse $\vcVar_1$ for both subdomains separately.
For a size $\sizeLow < \leftRangeBound + \rightRangeBound$, we get
$$
  \vcVar_1(\sizeLow)
  \quad \equiv \quad
  \forall \indexVar.\
    \slFalse \rightarrow (0 \leq \indexVar + \zVar < \sizeLow)
  \quad \equiv \quad
    \slTrue
$$
So, we do not have to bother checking sizes $\sizeVar < \leftRangeBound + \rightRangeBound$. For bigger sizes $\sizeHigh \geq \leftRangeBound + \rightRangeBound$, we get
$$
\begin{array}{l}
  \vcVar_1(\sizeHigh)
  \ \ \equiv\ \
  \forall \indexVar.\
      (\leftRangeBound \leq \indexVar \leq \sizeHigh - \rightRangeBound)
      \rightarrow
      (0 \leq \indexVar + \zVar < \sizeHigh)
  \\
  \equiv\ \
  \forall \indexVar.\
    (
      \leftRangeBound \leq \indexVar 
      \rightarrow 
      0 \leq \indexVar + \zVar
    )
    \ \wedge \
    (
      \indexVar \leq \sizeHigh - \rightRangeBound
      \rightarrow
      \indexVar + \zVar < \sizeHigh
    )
  \\
  \equiv\ \
  \forall \indexVar.\
    (
      \leftRangeBound \leq \indexVar 
      \rightarrow 
      0 \leq \indexVar + \zVar
    )
    \ \wedge \
    (
      \indexVar \leq  -\rightRangeBound
      \rightarrow
      \indexVar + \zVar < 0
    )
  \\
  =:\ \
  \vcVar_2.
\end{array}
$$

Since $\sizeHigh$ does not occur freely in $\vcVar_2$, the truth of $\vcVar_1(\sizeHigh)$ does not depend on the choice of $\sizeHigh$.
Remember that we have,
$\vcTrav(\sizeLow) \equiv \slTrue$
and
$\vcTrav(\sizeHigh) \equiv \vcVar_2$.
Hence,
$$
  \models 
    \forall \arrayVar.\ \forall \sizeVar.\ 
    \vcTrav
  \quad\Leftrightarrow\quad
  \models
    \forall \arrayVar.\
      \vcTrav(\sizeHigh)
$$
We see that it suffices to check the original VC $\vcTrav$ for any size $\sizeHigh \geq \leftRangeBound - \rightRangeBound$ to prove memory safety of our array traversing program \travPattern.
That is, $\{\sizeHigh\}$ is a CT.

\paragraph{Characterizing CTs via Constraints}
Note that the constraint
$\sizeVar \geq \leftRangeBound - \rightRangeBound$
we just derived
is a uniform representation for the CTs of the entire class \travPatternClass.
We often use constraints to concisely characterize CTs.
A constraint formulates a property that is sufficiently strong so that every subdomain covering it is a CT.
For every program $\travPattern \in \travPatternClass$, every subdomain $\ctVar' \subseteq \N$ is a CT for the array size \sizeVar if
$
  \ctVar' \cap 
  \{
    \sizeVar \in \N \ | \ 
    \sizeVar \geq \leftRangeBound - \rightRangeBound  
  \}
  \neq \emptyset
$.

\subsection{Modularity of Completeness Thresholds}
\label{sec:CTs:Modularity}

\paragraph{Unrelated Data Structures}
Consider the program\\
\sumPatternVars{\arrayVar, \sizeVar, \nVar} that attempts to sum up all elements of array~\arrayVar and writes the result to heap location~\nVar:

\begin{tabular}{lll}
  $\left.
    \begin{minipage}{4cm}
\begin{lstlisting}
for i in [L : s-R] do 
  !n := !n + !a[i+Z]
\end{lstlisting}
    \end{minipage}
  \right\}$
  &=:
  &\sumPatternVars{\arrayVar, \sizeVar, \nVar}
\end{tabular}
\\
Analogously to \travPattern it iterates through the array and attempts to read array elements.
Additionally, it uses the read value to update the sum stored at heap location \nVar.
Let us assume that the array \arrayVar and the result variable \nVar do not alias.
We get the specification
\hoareTriple{\preVar}{\sumPattern}{\preVar}
for 
$\preVar := \arrayPred \slStar \slPointsToAny{\nVar}$.
We assume structured memory.
Therefore, it is not possible to access \nVar via an array access~$\arrayVar[...]$.

Intuitively, it is clear that the array size does not affect the memory accesses to heap location~\nVar.
Hence, the CTs for~\sizeVar in~\sumPattern should be the same as the ones for~\travPattern.
In fact, analysing the \wlpName-based VC
for \sumPattern confirms this intuition.
It has the form 
$\forall \arrayVar. \forall \sizeVar. \forall \resVar.\ \vcSum$
and we can rewrite \vcSum into
$$
\begin{array}{l l l}
  \vcSum
  &\equiv
  &\vcTrav
    \slStar
    (\slPointsToAny{\nVar} \rightarrow \subAssVar)
\end{array}
$$
where \vcTrav is the VC from \S~\ref{sec:CTs:ApproxCTsViaVCs} for the array traversing program \travPattern.
Moreover, $\freeVars{\subAssVar} = \{\nVar\}$.
Since \sizeVar does not occur freely in
$\slPointsToAny{\nVar} \rightarrow \subAssVar$,
we can ignore it while searching for a CT for \sizeVar in \vcSum.

\begin{lemma}[VC Slicing]\label{lem:vcSlicing}
  Let $\assVar, \assVar_x, \assVar_y$ be assertions with 
  $x \in \freeVars{\assVar_x}$ 
  and 
  $x \not\in \freeVars{\assVar_y}$ 
  and 
  $\assVar \equiv \assVar_x \slStar \assVar_y$.
  Let $\ctVar \subseteq \domVar$ be a CT for $x$ in $\assVar_x$.
  Then, \ctVar is also a CT for $x$ in $\assVar$, i.e.,
  $$
    \models
      \forall x \in \ctVar.\ 
      \forall \overline{y} \in \overline{Y}.\
        \assVar
    \quad\Rightarrow\quad
      \models
      \forall x \in \domVar.\ 
      \forall \overline{y} \in \overline{Y}.\
        \assVar
  $$
\end{lemma}

\begin{figure}
  \raggedright
  \begin{tabular}{l l l}
    $\left.
    \begin{minipage}{4cm}
\begin{lstlisting}
for i in [L : s-R] do (
  n := a[i+Z];
  complex_fct(n, y, k)
)
\end{lstlisting}
      \end{minipage}
    \quad\right\}$
    &=:
    &\compProgVars{\arrayVar, \sizeVar, \unrelStructVar, \unrelSizeVar} 
  \end{tabular}
  
  \begin{tabular}{l l l}
    $\memAssVar(\arrayVar, \sizeVar)$
    &:=
    &\arrayPred \slStar\ \unrelDataPred
    \\
    \compProgClass
    &:=
    &$\{ 
      \hoareTriple{\memAssVar}{\compProg}{\memAssVar}
      \quad | \quad 
      \leftRangeBound, \rightRangeBound, \zVar \in \Z 
    \}$
  \end{tabular}

  \caption{
    Class of programs involving a complex data structure and computation that do not depend on the array size~\sizeVar.
  }
  \label{fig:ComplexProgClass}
\end{figure}

We can extrapolate what we saw in the \sumPattern example to more complex classes of programs.
Consider the class~\compProgClass presented in Fig.~\ref{fig:ComplexProgClass}.
A program $\compProg \in \compProgClass$ receives two non-aliasing data structures: an array \arrayVar of size \sizeVar and a complex data structure \unrelStructVar of size \unrelSizeVar, described by the predicate \unrelDataPred.
\compProg reads elements from array \arrayVar, stores the result in a local variable \nVar and then calls a complex function
\verb|complex_fct(n,y,k)|
which does neither depend on \arrayVar nor \sizeVar.
The VC \vcComp will reflect this.
That is, analogous to the example above, it should be expressible as
$
  \vcComp 
  \equiv 
    \vcTrav
    \slStar 
    (\unrelDataPred \rightarrow \dots)
$
where the right conjunct does not depend on \sizeVar.

VC Slicing lemma~\ref{lem:vcSlicing} tells us that whenever we want to characterize a CT for a specific parameter, 
we can ignore all separated VC conjuncts that do not involve this parameter.
Effectively, this means that we can ignore all the complex parts of \compProg that are not related to the array size while searching for a CT for \sizeVar.
This allows us to reduce the search to the CTs of~\travPatternClass.

\paragraph{Compositionality}
We can describe the CTs of complex programs in terms of the CTs of their building blocks.
Consider the program $\cmdVar_1;\cmdVar_2$ and suppose that $\cmdVar_1$ and $\cmdVar_2$ are instances of patterns we studied before.
So we know that each $\cmdVar_i$ corresponds to a VC 
$\forall x.\ \vcVar_i$ 
with a CT $\ctVar_i$ for $x$.
Let $\vcVar_{1;2}$ be the VC for $\cmdVar_1; \cmdVar_2$ that we want to prove.
Suppose it can be rewritten into 
$\forall x.\ \vcVar_1 \wedge \vcVar_2$.
Then, we know that $\ctVar_1 \cup \ctVar_2$ is a CT for $\vcVar_{1;2}$.
Therefore, our approach to studying CTs is to study patterns and combinators.

\paragraph{Basic Patterns}
We view basic patterns such as the array traversal pattern discussed above as the basic building blocks.
They tend to occur frequently in programs and they are sufficiently concise to extract CTs by studying their VCs.
In particular, we focus on traversal and access patterns that preserve the memory layout.
For now, we focus on arrays, but we are going to generalize it to arbitrary inductive data structures.

\paragraph{Managing Complexity}
One of our main goals is to describe CTs for interesting classes of programs.
VCs tend to become very complex very fast as a program gets more complex.
Hence, we need a way to deal with this complexity and to break the CT analysis down into simpler problems.
Following the structure of the program we want to reason about is a natural approach.

\paragraph{Combinator Patterns}
In order to exploit the program structure while analysing CTs, we need to study how control structures affect CTs.
For instance, as described above, we can characterize the CT of a sequence $\cmdVar_1; \cmdVar_2$ as the union of the CTs derived from $\cmdVar_1$ and $\cmdVar_2$.
Further, consider the command 
$\texttt{if}\ e\ \texttt{then}\ \cmdVar_1\ \texttt{else}\ \cmdVar_2$
and suppose that we can describe CTs for a size \sizeVar in each $\cmdVar_i$ via constraints $\bigwedge\constraintSetVar_i$.
Then, we can describe the CT for the entire command via the constraints
$
  (e \wedge \bigwedge K_1)
  \ \vee\
  (\neg e \wedge \bigwedge K_2)
$.

	\section{Conclusion}\label{sec:Conclusion}

Past approaches to program verification either targeted unbounded guarantees and relied on unbounded, often inductive, proofs or they targeted bounded guarantees and tried to approximate the program behaviour using techniques like bounded model checking.
We have, however, seen little interaction between the two communities.

In this work we propose a new perspective on memory safety proofs that connects unbounded and bounded proofs.
We show that we can reduce unbounded memory safety proofs to bounded ones for certain programs that traverse arrays and preserve the memory layout.
For any such program considering a few select array sizes yields the same guarantees as considering arrays of all possible sizes.
We call this concept \emph{completeness thresholds} in reference to a similar concept from model checking of finite transition systems.
Moreover, we show that studying verification conditions are an adequate way to study completeness thresholds.

	\section{Related Work}\label{sec:RelatedWork}

Completeness thresholds were first introduced by \citeauthor{Kroening2003EfficientCO}~\cite{Kroening2003EfficientCO}.
So far, the study of CTs has been limited to finite state systems.
Indeed, the well-known CTs for classes of LTL properties are defined with respect to the (recurrence) diameter of the finite state system in question (e.g.\ \cite{Biere1999SymbolicMC,DBLP:conf/cav/KroeningOSWW11}).
Determining the worst-case execution time of a program and discovering upper bounds on loops by iterative unrolling can also be used to determine CTs~\cite{DBLP:journals/tcad/DSilvaKW08,DBLP:conf/tacas/ClarkeKL04}.
For a possibly infinite state system those CTs can naturally be infinite as well.
By specializing in just one property, memory safety, we are able to characterize and possibly find useful CTs for these systems as well.
Model checking for parameterized network topologies of identical (e.g.\ bisimilar~\cite{DBLP:journals/iandc/ClarkeGB89} or isomorphic~\cite{DBLP:conf/popl/EmersonN95}) processes features a related concept to completeness thresholds called \emph{cutoff}.
That is, model-checking up to the cutoff implies correctness of scaling the topology up to infinitely many processes.
Positive results exist for properties of such token rings~\cite{DBLP:journals/iandc/ClarkeGB89,DBLP:conf/popl/EmersonN95} but also other topologies~\cite{DBLP:conf/concur/ClarkeTTV04,DBLP:conf/vmcai/AminofJKR14}.

The model checking literature (cf.~\cite{DBLP:books/daglib/0007403-2}) boasts a wealth of alternative approaches to obtain unbounded guarantees on finite state systems, e.g.,
k-induction~\cite{DBLP:conf/fmcad/BjesseC00,DBLP:conf/fmcad/SheeranSS00}, 
Craig interpolation~\cite{McMillan2003InterpolationAS} and 
property-directed reachability~\cite{DBLP:conf/vmcai/Bradley11,DBLP:conf/fmcad/EenMB11}
with adaptions to the software verification setting (e.g.~\cite{DBLP:conf/sas/DonaldsonHKR11,DBLP:journals/corr/abs-2208-05046,DBLP:conf/tacas/0001D20}).

Array-manipulating programs are well-studied across different domains~\cite{DBLP:conf/cav/BozgaHIKV09, Bradley2006DecidableArrays, JhalaM2007ArrayAbstraction,
Chakraborty2020ArrayFullInduction}.
However, we consider our main contribution to be a novel approach to connect unbounded and bounded proofs about memory safety.
Ultimately, as discussed in the outlook, we aim to generalize and automate our approach to tree-like data structures.
In that regard, \citeauthor{DBLP:journals/pacmpl/MathurMKMV20}~\cite{DBLP:journals/pacmpl/MathurMKMV20} consider the special case of proving memory safety of heap-manipulating programs as well.
They prove that memory safety is decidable if the initial heap is forest-like and the program only performs a single-pass over the data-structure (see also \S~\ref{sec:Outlook}).
%
%
They do not cover arrays and buffer overflows, but they support (de-)allocation.

	\section{Outlook}\label{sec:Outlook}

\paragraph{CTs for Programs}
Consider any program \cmdVar with a free variable $x$ and a corresponding VC \vcVar.
Suppose we derived a CT \ctVar for $x$ in \vcVar.
In general, this does not allow us to conclude that \ctVar is also a CT for $x$ in \cmdVar.
Intuitively, this is only true if \vcVar does not over-approximate with regard to $x$.
This holds for the programs and VCs studied in this paper.
While CTs for VCs still tell us something about proofs targeting these VCs, our ultimate goal is to derive thresholds for programs.
Hence, we are currently studying this connection.

\paragraph{Scalability}
In this work, we focus on a restricted array traversal pattern to illustrate CTs.
Our goal is to scale this approach to complex programs.
Therefore, we are currently studying more array traversal and access patterns and combinators.
This will allow us to better understand how the structure of programs affects the relation between bounded and unbounded proofs.
Knowing this will allow us to characterise CTs for complex classes of programs that cover errors besides off-by-n errors.

Afterwards, we are going to extend our approach to include arbitrary inductive data types.
In particular, we plan to describe CTs for a class of programs $\mathit{Sort}$ that includes (safe and unsafe) implementations of in-place sorting algorithms involving nested loops.
Once we managed that, we are going to investigate how allocation and deallocation affect CTs.

\paragraph{Decidability}
The memory accesses we observe in sorting functions typically mainly depend on the size of the sorted data structure, the traversal strategy and a comparison relation $<$.
Suppose we managed to derive finite CTs for the class $\mathit{Sort}$.
We conjecture that this will be sufficient to conclude decidability of memory safety for $\mathit{Sort}$.
Note that this targeted result would escape the scope of the related work by 
\citeauthor{DBLP:journals/pacmpl/MathurMKMV20}~\cite{DBLP:journals/pacmpl/MathurMKMV20}.
The latter showed decidability of memory safety for a certain form of single-pass programs, i.e., programs that traverse a datastructure exactly once.

\paragraph{Improving BMC Guarantees}
Bounded model checking suffers from the state space explosion problem~\cite{Clarke2008BmcStateExplosion, Park2000JavaMC, Clarke2000CounterExGuidedRefinement}
with respect to the chosen bounds.
It is often only practical to check very small bounds on each parameter to keep the verification time practical.
We plan to automate the approach introduced in this paper and to leverage existing static analysis techniques (e.g.~\cite{Weiser84ProgramSlicing, MastroeniDataDependencyProgramSlicing, Asavoae2018ChiselProgramSlicing, Moser90DataDependencyGraphs}) to simplify the generated VCs. 
Once we are able to automatically compute a small CT for one parameter, say the traversed data structure's size, we can adjust the corresponding bound.
We can be sure that we didn't miss to check any size that could lead to memory errors.
Hence, it would strengthen the guarantees we get from our bounded proof.
Ultimately, we would like to integrate
 with the industrial-strength BMC tool\\ CBMC~\cite{DBLP:conf/dac/ClarkeKY03,DBLP:conf/tacas/ClarkeKL04}.\phantom{aaaabbbbccccdddd}

Further, it would speed up the verification time and free resources that can be spend on exploring other parts of the program.
That is, the lowered size bound might allow us to increase other bounds that seem more important.
We would be able to do this while keeping the verification time stable and without sacrificing guarantees.

\balance

	\bibliographystyle{ACM-Reference-Format}
	\bibliography{bibliography}


\begin{thebibliography}{47}


\ifx \showCODEN    \undefined \def \showCODEN     #1{\unskip}     \fi
\ifx \showDOI      \undefined \def \showDOI       #1{#1}\fi
\ifx \showISBNx    \undefined \def \showISBNx     #1{\unskip}     \fi
\ifx \showISBNxiii \undefined \def \showISBNxiii  #1{\unskip}     \fi
\ifx \showISSN     \undefined \def \showISSN      #1{\unskip}     \fi
\ifx \showLCCN     \undefined \def \showLCCN      #1{\unskip}     \fi
\ifx \shownote     \undefined \def \shownote      #1{#1}          \fi
\ifx \showarticletitle \undefined \def \showarticletitle #1{#1}   \fi
\ifx \showURL      \undefined \def \showURL       {\relax}        \fi
\providecommand\bibfield[2]{#2}
\providecommand\bibinfo[2]{#2}
\providecommand\natexlab[1]{#1}
\providecommand\showeprint[2][]{arXiv:#2}

\bibitem[Abdulaziz et~al\mbox{.}(2018)]%
        {Abdulaziz2018FormallyVA}
\bibfield{author}{\bibinfo{person}{Mohammad Abdulaziz},
  \bibinfo{person}{Michael Norrish}, {and} \bibinfo{person}{Charles Gretton}.}
  \bibinfo{year}{2018}\natexlab{}.
\newblock \showarticletitle{Formally Verified Algorithms for Upper-Bounding
  State Space Diameters}.
\newblock \bibinfo{journal}{\emph{Journal of Automated Reasoning}}
  \bibinfo{volume}{61} (\bibinfo{year}{2018}), \bibinfo{pages}{485--520}.
\newblock


\bibitem[Aminof et~al\mbox{.}(2014)]%
        {DBLP:conf/vmcai/AminofJKR14}
\bibfield{author}{\bibinfo{person}{Benjamin Aminof}, \bibinfo{person}{Swen
  Jacobs}, \bibinfo{person}{Ayrat Khalimov}, {and} \bibinfo{person}{Sasha
  Rubin}.} \bibinfo{year}{2014}\natexlab{}.
\newblock \showarticletitle{Parameterized Model Checking of Token-Passing
  Systems}. In \bibinfo{booktitle}{\emph{Verification, Model Checking, and
  Abstract Interpretation - 15th International Conference, {VMCAI} 2014, San
  Diego, CA, USA, January 19-21, 2014, Proceedings}}
  \emph{(\bibinfo{series}{Lecture Notes in Computer Science},
  Vol.~\bibinfo{volume}{8318})}, \bibfield{editor}{\bibinfo{person}{Kenneth~L.
  McMillan} {and} \bibinfo{person}{Xavier Rival}} (Eds.).
  \bibinfo{publisher}{Springer}, \bibinfo{pages}{262--281}.
\newblock
\urldef\tempurl%
\url{https://doi.org/10.1007/978-3-642-54013-4\_15}
\showDOI{\tempurl}


\bibitem[Asavoae et~al\mbox{.}(2018)]%
        {Asavoae2018ChiselProgramSlicing}
\bibfield{author}{\bibinfo{person}{Irina~Mariuca Asavoae},
  \bibinfo{person}{Mihail Asavoae}, {and} \bibinfo{person}{Adri{\'{a}}n
  Riesco}.} \bibinfo{year}{2018}\natexlab{}.
\newblock \showarticletitle{Slicing from formal semantics: Chisel - a tool for
  generic program slicing}.
\newblock \bibinfo{journal}{\emph{Int. J. Softw. Tools Technol. Transf.}}
  \bibinfo{volume}{20}, \bibinfo{number}{6} (\bibinfo{year}{2018}),
  \bibinfo{pages}{739--769}.
\newblock
\urldef\tempurl%
\url{https://doi.org/10.1007/s10009-018-0500-y}
\showDOI{\tempurl}


\bibitem[Awedh and Somenzi(2004)]%
        {Awedh2004ProvingMP}
\bibfield{author}{\bibinfo{person}{Mohammad Awedh} {and} \bibinfo{person}{F.
  Somenzi}.} \bibinfo{year}{2004}\natexlab{}.
\newblock \showarticletitle{Proving More Properties with Bounded Model
  Checking}. In \bibinfo{booktitle}{\emph{CAV}}.
\newblock


\bibitem[Beyer and Dangl(2020)]%
        {DBLP:conf/tacas/0001D20}
\bibfield{author}{\bibinfo{person}{Dirk Beyer} {and} \bibinfo{person}{Matthias
  Dangl}.} \bibinfo{year}{2020}\natexlab{}.
\newblock \showarticletitle{Software Verification with {PDR:} An Implementation
  of the State of the Art}. In \bibinfo{booktitle}{\emph{Tools and Algorithms
  for the Construction and Analysis of Systems - 26th International Conference,
  {TACAS} 2020, Held as Part of the European Joint Conferences on Theory and
  Practice of Software, {ETAPS} 2020, Dublin, Ireland, April 25-30, 2020,
  Proceedings, Part {I}}} \emph{(\bibinfo{series}{Lecture Notes in Computer
  Science}, Vol.~\bibinfo{volume}{12078})},
  \bibfield{editor}{\bibinfo{person}{Armin Biere} {and} \bibinfo{person}{David
  Parker}} (Eds.). \bibinfo{publisher}{Springer}, \bibinfo{pages}{3--21}.
\newblock
\urldef\tempurl%
\url{https://doi.org/10.1007/978-3-030-45190-5\_1}
\showDOI{\tempurl}


\bibitem[Beyer et~al\mbox{.}(2022)]%
        {DBLP:journals/corr/abs-2208-05046}
\bibfield{author}{\bibinfo{person}{Dirk Beyer}, \bibinfo{person}{Nian{-}Ze
  Lee}, {and} \bibinfo{person}{Philipp Wendler}.}
  \bibinfo{year}{2022}\natexlab{}.
\newblock \showarticletitle{Interpolation and SAT-Based Model Checking
  Revisited: Adoption to Software Verification}.
\newblock \bibinfo{journal}{\emph{CoRR}}  \bibinfo{volume}{abs/2208.05046}
  (\bibinfo{year}{2022}).
\newblock
\urldef\tempurl%
\url{https://doi.org/10.48550/arXiv.2208.05046}
\showDOI{\tempurl}
\showeprint[arXiv]{2208.05046}


\bibitem[Biere et~al\mbox{.}(1999)]%
        {Biere1999SymbolicMC}
\bibfield{author}{\bibinfo{person}{Armin Biere}, \bibinfo{person}{Alessandro
  Cimatti}, \bibinfo{person}{Edmund~M. Clarke}, {and} \bibinfo{person}{Yunshan
  Zhu}.} \bibinfo{year}{1999}\natexlab{}.
\newblock \showarticletitle{Symbolic Model Checking without BDDs}. In
  \bibinfo{booktitle}{\emph{International Conference on Tools and Algorithms
  for Construction and Analysis of Systems}}.
\newblock


\bibitem[Bjesse and Claessen(2000)]%
        {DBLP:conf/fmcad/BjesseC00}
\bibfield{author}{\bibinfo{person}{Per Bjesse} {and} \bibinfo{person}{Koen
  Claessen}.} \bibinfo{year}{2000}\natexlab{}.
\newblock \showarticletitle{SAT-Based Verification without State Space
  Traversal}. In \bibinfo{booktitle}{\emph{Formal Methods in Computer-Aided
  Design, Third International Conference, {FMCAD} 2000, Austin, Texas, USA,
  November 1-3, 2000, Proceedings}} \emph{(\bibinfo{series}{Lecture Notes in
  Computer Science}, Vol.~\bibinfo{volume}{1954})},
  \bibfield{editor}{\bibinfo{person}{Warren A.~Hunt Jr.} {and}
  \bibinfo{person}{Steven~D. Johnson}} (Eds.). \bibinfo{publisher}{Springer},
  \bibinfo{pages}{372--389}.
\newblock
\urldef\tempurl%
\url{https://doi.org/10.1007/3-540-40922-X\_23}
\showDOI{\tempurl}


\bibitem[Bozga et~al\mbox{.}(2009)]%
        {DBLP:conf/cav/BozgaHIKV09}
\bibfield{author}{\bibinfo{person}{Marius Bozga}, \bibinfo{person}{Peter
  Habermehl}, \bibinfo{person}{Radu Iosif}, \bibinfo{person}{Filip
  Konecn{\'{y}}}, {and} \bibinfo{person}{Tom{\'{a}}s Vojnar}.}
  \bibinfo{year}{2009}\natexlab{}.
\newblock \showarticletitle{Automatic Verification of Integer Array Programs}.
  In \bibinfo{booktitle}{\emph{Computer Aided Verification, 21st International
  Conference, {CAV} 2009, Grenoble, France, June 26 - July 2, 2009.
  Proceedings}} \emph{(\bibinfo{series}{Lecture Notes in Computer Science},
  Vol.~\bibinfo{volume}{5643})}, \bibfield{editor}{\bibinfo{person}{Ahmed
  Bouajjani} {and} \bibinfo{person}{Oded Maler}} (Eds.).
  \bibinfo{publisher}{Springer}, \bibinfo{pages}{157--172}.
\newblock
\urldef\tempurl%
\url{https://doi.org/10.1007/978-3-642-02658-4\_15}
\showDOI{\tempurl}


\bibitem[Bradley(2011)]%
        {DBLP:conf/vmcai/Bradley11}
\bibfield{author}{\bibinfo{person}{Aaron~R. Bradley}.}
  \bibinfo{year}{2011}\natexlab{}.
\newblock \showarticletitle{SAT-Based Model Checking without Unrolling}. In
  \bibinfo{booktitle}{\emph{Verification, Model Checking, and Abstract
  Interpretation - 12th International Conference, {VMCAI} 2011, Austin, TX,
  USA, January 23-25, 2011. Proceedings}} \emph{(\bibinfo{series}{Lecture Notes
  in Computer Science}, Vol.~\bibinfo{volume}{6538})},
  \bibfield{editor}{\bibinfo{person}{Ranjit Jhala} {and}
  \bibinfo{person}{David~A. Schmidt}} (Eds.). \bibinfo{publisher}{Springer},
  \bibinfo{pages}{70--87}.
\newblock
\urldef\tempurl%
\url{https://doi.org/10.1007/978-3-642-18275-4\_7}
\showDOI{\tempurl}


\bibitem[Bradley et~al\mbox{.}(2006)]%
        {Bradley2006DecidableArrays}
\bibfield{author}{\bibinfo{person}{Aaron~R. Bradley}, \bibinfo{person}{Zohar
  Manna}, {and} \bibinfo{person}{Henny~B. Sipma}.}
  \bibinfo{year}{2006}\natexlab{}.
\newblock \showarticletitle{What's Decidable About Arrays?}. In
  \bibinfo{booktitle}{\emph{Verification, Model Checking, and Abstract
  Interpretation, 7th International Conference, {VMCAI} 2006, Charleston, SC,
  USA, January 8-10, 2006, Proceedings}} \emph{(\bibinfo{series}{Lecture Notes
  in Computer Science}, Vol.~\bibinfo{volume}{3855})},
  \bibfield{editor}{\bibinfo{person}{E.~Allen Emerson} {and}
  \bibinfo{person}{Kedar~S. Namjoshi}} (Eds.). \bibinfo{publisher}{Springer},
  \bibinfo{pages}{427--442}.
\newblock
\urldef\tempurl%
\url{https://doi.org/10.1007/11609773\_28}
\showDOI{\tempurl}


\bibitem[Browne et~al\mbox{.}(1989)]%
        {DBLP:journals/iandc/ClarkeGB89}
\bibfield{author}{\bibinfo{person}{Michael~C. Browne},
  \bibinfo{person}{Edmund~M. Clarke}, {and} \bibinfo{person}{Orna Grumberg}.}
  \bibinfo{year}{1989}\natexlab{}.
\newblock \showarticletitle{Reasoning about Networks with Many Identical Finite
  State Processes}.
\newblock \bibinfo{journal}{\emph{Inf. Comput.}} \bibinfo{volume}{81},
  \bibinfo{number}{1} (\bibinfo{year}{1989}), \bibinfo{pages}{13--31}.
\newblock
\urldef\tempurl%
\url{https://doi.org/10.1016/0890-5401(89)90026-6}
\showDOI{\tempurl}


\bibitem[Bundala et~al\mbox{.}(2012)]%
        {Bundala2012OnTM}
\bibfield{author}{\bibinfo{person}{Daniel Bundala}, \bibinfo{person}{Jo{\"e}l
  Ouaknine}, {and} \bibinfo{person}{James Worrell}.}
  \bibinfo{year}{2012}\natexlab{}.
\newblock \showarticletitle{On the Magnitude of Completeness Thresholds in
  Bounded Model Checking}.
\newblock \bibinfo{journal}{\emph{27th Annual IEEE Symposium on Logic in
  Computer Science}} (\bibinfo{year}{2012}), \bibinfo{pages}{155--164}.
\newblock


\bibitem[Chakraborty et~al\mbox{.}(2020)]%
        {Chakraborty2020ArrayFullInduction}
\bibfield{author}{\bibinfo{person}{Supratik Chakraborty},
  \bibinfo{person}{Ashutosh Gupta}, {and} \bibinfo{person}{Divyesh Unadkat}.}
  \bibinfo{year}{2020}\natexlab{}.
\newblock \showarticletitle{Verifying Array Manipulating Programs with
  Full-Program Induction}. In \bibinfo{booktitle}{\emph{Tools and Algorithms
  for the Construction and Analysis of Systems - 26th International Conference,
  {TACAS} 2020, Held as Part of the European Joint Conferences on Theory and
  Practice of Software, {ETAPS} 2020, Dublin, Ireland, April 25-30, 2020,
  Proceedings, Part {I}}} \emph{(\bibinfo{series}{Lecture Notes in Computer
  Science}, Vol.~\bibinfo{volume}{12078})},
  \bibfield{editor}{\bibinfo{person}{Armin Biere} {and} \bibinfo{person}{David
  Parker}} (Eds.). \bibinfo{publisher}{Springer}, \bibinfo{pages}{22--39}.
\newblock
\urldef\tempurl%
\url{https://doi.org/10.1007/978-3-030-45190-5\_2}
\showDOI{\tempurl}


\bibitem[Chong et~al\mbox{.}(2020)]%
        {Chong2020CodeLeveMC}
\bibfield{author}{\bibinfo{person}{Nathan Chong}, \bibinfo{person}{Byron Cook},
  \bibinfo{person}{Konstantinos Kallas}, \bibinfo{person}{Kareem Khazem},
  \bibinfo{person}{Felipe~R. Monteiro}, \bibinfo{person}{Daniel
  Schwartz-Narbonne}, \bibinfo{person}{Serdar Tasiran},
  \bibinfo{person}{Michael Tautschnig}, {and} \bibinfo{person}{Mark~R.
  Tuttle}.} \bibinfo{year}{2020}\natexlab{}.
\newblock \showarticletitle{Code-Level Model Checking in the Software
  Development Workflow}. In \bibinfo{booktitle}{\emph{Proceedings of the
  ACM/IEEE 42nd International Conference on Software Engineering: Software
  Engineering in Practice}} (Seoul, South Korea)
  \emph{(\bibinfo{series}{ICSE-SEIP '20})}. \bibinfo{publisher}{Association for
  Computing Machinery}, \bibinfo{address}{New York, NY, USA},
  \bibinfo{pages}{11–20}.
\newblock
\showISBNx{9781450371230}
\urldef\tempurl%
\url{https://doi.org/10.1145/3377813.3381347}
\showDOI{\tempurl}


\bibitem[Clarke et~al\mbox{.}(2000)]%
        {Clarke2000CounterExGuidedRefinement}
\bibfield{author}{\bibinfo{person}{Edmund Clarke}, \bibinfo{person}{Orna
  Grumberg}, \bibinfo{person}{Somesh Jha}, \bibinfo{person}{Yuan Lu}, {and}
  \bibinfo{person}{Helmut Veith}.} \bibinfo{year}{2000}\natexlab{}.
\newblock \showarticletitle{Counterexample-Guided Abstraction Refinement}. In
  \bibinfo{booktitle}{\emph{Computer Aided Verification}},
  \bibfield{editor}{\bibinfo{person}{E.~Allen Emerson} {and}
  \bibinfo{person}{Aravinda~Prasad Sistla}} (Eds.).
  \bibinfo{publisher}{Springer Berlin Heidelberg}, \bibinfo{address}{Berlin,
  Heidelberg}, \bibinfo{pages}{154--169}.
\newblock


\bibitem[Clarke(2008)]%
        {Clarke2008BmcStateExplosion}
\bibfield{author}{\bibinfo{person}{Edmund~M. Clarke}.}
  \bibinfo{year}{2008}\natexlab{}.
\newblock \showarticletitle{Model Checking -- My 27-Year Quest to Overcome the
  State Explosion Problem}. In \bibinfo{booktitle}{\emph{Logic for Programming,
  Artificial Intelligence, and Reasoning}},
  \bibfield{editor}{\bibinfo{person}{Iliano Cervesato}, \bibinfo{person}{Helmut
  Veith}, {and} \bibinfo{person}{Andrei Voronkov}} (Eds.).
  \bibinfo{publisher}{Springer Berlin Heidelberg}, \bibinfo{address}{Berlin,
  Heidelberg}, \bibinfo{pages}{182--182}.
\newblock
\showISBNx{978-3-540-89439-1}


\bibitem[Clarke et~al\mbox{.}(2018)]%
        {DBLP:books/daglib/0007403-2}
\bibfield{author}{\bibinfo{person}{Edmund~M. Clarke}, \bibinfo{person}{Orna
  Grumberg}, \bibinfo{person}{Daniel Kroening}, \bibinfo{person}{Doron~A.
  Peled}, {and} \bibinfo{person}{Helmut Veith}.}
  \bibinfo{year}{2018}\natexlab{}.
\newblock \bibinfo{booktitle}{\emph{Model checking, 2nd Edition}}.
\newblock \bibinfo{publisher}{{MIT} Press}.
\newblock
\showISBNx{978-0-262-03883-6}
\urldef\tempurl%
\url{https://mitpress.mit.edu/books/model-checking-second-edition}
\showURL{%
\tempurl}


\bibitem[Clarke et~al\mbox{.}(2004a)]%
        {DBLP:conf/tacas/ClarkeKL04}
\bibfield{author}{\bibinfo{person}{Edmund~M. Clarke}, \bibinfo{person}{Daniel
  Kroening}, {and} \bibinfo{person}{Flavio Lerda}.}
  \bibinfo{year}{2004}\natexlab{a}.
\newblock \showarticletitle{A Tool for Checking {ANSI-C} Programs}. In
  \bibinfo{booktitle}{\emph{Tools and Algorithms for the Construction and
  Analysis of Systems, 10th International Conference, {TACAS} 2004, Held as
  Part of the Joint European Conferences on Theory and Practice of Software,
  {ETAPS} 2004, Barcelona, Spain, March 29 - April 2, 2004, Proceedings}}
  \emph{(\bibinfo{series}{Lecture Notes in Computer Science},
  Vol.~\bibinfo{volume}{2988})}, \bibfield{editor}{\bibinfo{person}{Kurt
  Jensen} {and} \bibinfo{person}{Andreas Podelski}} (Eds.).
  \bibinfo{publisher}{Springer}, \bibinfo{pages}{168--176}.
\newblock
\urldef\tempurl%
\url{https://doi.org/10.1007/978-3-540-24730-2\_15}
\showDOI{\tempurl}


\bibitem[Clarke et~al\mbox{.}(2004b)]%
        {Clarke2004CompletenessAC}
\bibfield{author}{\bibinfo{person}{Edmund~M. Clarke}, \bibinfo{person}{Daniel
  Kroening}, \bibinfo{person}{Jo{\"e}l Ouaknine}, {and} \bibinfo{person}{Ofer
  Strichman}.} \bibinfo{year}{2004}\natexlab{b}.
\newblock \showarticletitle{Completeness and Complexity of Bounded Model
  Checking}. In \bibinfo{booktitle}{\emph{International Conference on
  Verification, Model Checking and Abstract Interpretation}}.
\newblock


\bibitem[Clarke et~al\mbox{.}(2003)]%
        {DBLP:conf/dac/ClarkeKY03}
\bibfield{author}{\bibinfo{person}{Edmund~M. Clarke}, \bibinfo{person}{Daniel
  Kroening}, {and} \bibinfo{person}{Karen Yorav}.}
  \bibinfo{year}{2003}\natexlab{}.
\newblock \showarticletitle{Behavioral consistency of {C} and verilog programs
  using bounded model checking}. In \bibinfo{booktitle}{\emph{Proceedings of
  the 40th Design Automation Conference, {DAC} 2003, Anaheim, CA, USA, June
  2-6, 2003}}. \bibinfo{publisher}{{ACM}}, \bibinfo{pages}{368--371}.
\newblock
\urldef\tempurl%
\url{https://doi.org/10.1145/775832.775928}
\showDOI{\tempurl}


\bibitem[Clarke et~al\mbox{.}(2004c)]%
        {DBLP:conf/concur/ClarkeTTV04}
\bibfield{author}{\bibinfo{person}{Edmund~M. Clarke},
  \bibinfo{person}{Muralidhar Talupur}, \bibinfo{person}{Tayssir Touili}, {and}
  \bibinfo{person}{Helmut Veith}.} \bibinfo{year}{2004}\natexlab{c}.
\newblock \showarticletitle{Verification by Network Decomposition}. In
  \bibinfo{booktitle}{\emph{{CONCUR} 2004 - Concurrency Theory, 15th
  International Conference, London, UK, August 31 - September 3, 2004,
  Proceedings}} \emph{(\bibinfo{series}{Lecture Notes in Computer Science},
  Vol.~\bibinfo{volume}{3170})}, \bibfield{editor}{\bibinfo{person}{Philippa
  Gardner} {and} \bibinfo{person}{Nobuko Yoshida}} (Eds.).
  \bibinfo{publisher}{Springer}, \bibinfo{pages}{276--291}.
\newblock
\urldef\tempurl%
\url{https://doi.org/10.1007/978-3-540-28644-8\_18}
\showDOI{\tempurl}


\bibitem[Corporation(2006)]%
        {cwe193}
\bibfield{author}{\bibinfo{person}{The~MITRE Corporation}.}
  \bibinfo{year}{2006}\natexlab{}.
\newblock \bibinfo{title}{CWE-193: Off-by-one Error}.
\newblock
\newblock
\urldef\tempurl%
\url{https://cwe.mitre.org/data/definitions/193.html}
\showURL{%
\tempurl}


\bibitem[Dijkstra(1976)]%
        {Dijsktra1976DisciplineOfProgramming}
\bibfield{author}{\bibinfo{person}{Edsger~W. Dijkstra}.}
  \bibinfo{year}{1976}\natexlab{}.
\newblock \showarticletitle{A Discipline of Programming}.
  \bibinfo{publisher}{Pentice Hall}.
\newblock


\bibitem[Donaldson et~al\mbox{.}(2011)]%
        {DBLP:conf/sas/DonaldsonHKR11}
\bibfield{author}{\bibinfo{person}{Alastair~F. Donaldson},
  \bibinfo{person}{Leopold Haller}, \bibinfo{person}{Daniel Kroening}, {and}
  \bibinfo{person}{Philipp R{\"{u}}mmer}.} \bibinfo{year}{2011}\natexlab{}.
\newblock \showarticletitle{Software Verification Using k-Induction}. In
  \bibinfo{booktitle}{\emph{Static Analysis - 18th International Symposium,
  {SAS} 2011, Venice, Italy, September 14-16, 2011. Proceedings}}
  \emph{(\bibinfo{series}{Lecture Notes in Computer Science},
  Vol.~\bibinfo{volume}{6887})}, \bibfield{editor}{\bibinfo{person}{Eran
  Yahav}} (Ed.). \bibinfo{publisher}{Springer}, \bibinfo{pages}{351--368}.
\newblock
\urldef\tempurl%
\url{https://doi.org/10.1007/978-3-642-23702-7\_26}
\showDOI{\tempurl}


\bibitem[D'Silva et~al\mbox{.}(2008)]%
        {DBLP:journals/tcad/DSilvaKW08}
\bibfield{author}{\bibinfo{person}{Vijay~Victor D'Silva},
  \bibinfo{person}{Daniel Kroening}, {and} \bibinfo{person}{Georg
  Weissenbacher}.} \bibinfo{year}{2008}\natexlab{}.
\newblock \showarticletitle{A Survey of Automated Techniques for Formal
  Software Verification}.
\newblock \bibinfo{journal}{\emph{{IEEE} Trans. Comput. Aided Des. Integr.
  Circuits Syst.}} \bibinfo{volume}{27}, \bibinfo{number}{7}
  (\bibinfo{year}{2008}), \bibinfo{pages}{1165--1178}.
\newblock
\urldef\tempurl%
\url{https://doi.org/10.1109/TCAD.2008.923410}
\showDOI{\tempurl}


\bibitem[E{\'{e}}n et~al\mbox{.}(2011)]%
        {DBLP:conf/fmcad/EenMB11}
\bibfield{author}{\bibinfo{person}{Niklas E{\'{e}}n}, \bibinfo{person}{Alan
  Mishchenko}, {and} \bibinfo{person}{Robert~K. Brayton}.}
  \bibinfo{year}{2011}\natexlab{}.
\newblock \showarticletitle{Efficient implementation of property directed
  reachability}. In \bibinfo{booktitle}{\emph{International Conference on
  Formal Methods in Computer-Aided Design, {FMCAD} '11, Austin, TX, USA,
  October 30 - November 02, 2011}}, \bibfield{editor}{\bibinfo{person}{Per
  Bjesse} {and} \bibinfo{person}{Anna Slobodov{\'{a}}}} (Eds.).
  \bibinfo{publisher}{{FMCAD} Inc.}, \bibinfo{pages}{125--134}.
\newblock
\urldef\tempurl%
\url{http://dl.acm.org/citation.cfm?id=2157675}
\showURL{%
\tempurl}


\bibitem[Emerson and Namjoshi(1995)]%
        {DBLP:conf/popl/EmersonN95}
\bibfield{author}{\bibinfo{person}{E.~Allen Emerson} {and}
  \bibinfo{person}{Kedar~S. Namjoshi}.} \bibinfo{year}{1995}\natexlab{}.
\newblock \showarticletitle{Reasoning about Rings}. In
  \bibinfo{booktitle}{\emph{Conference Record of POPL'95: 22nd {ACM}
  {SIGPLAN-SIGACT} Symposium on Principles of Programming Languages, San
  Francisco, California, USA, January 23-25, 1995}},
  \bibfield{editor}{\bibinfo{person}{Ron~K. Cytron} {and}
  \bibinfo{person}{Peter Lee}} (Eds.). \bibinfo{publisher}{{ACM} Press},
  \bibinfo{pages}{85--94}.
\newblock
\urldef\tempurl%
\url{https://doi.org/10.1145/199448.199468}
\showDOI{\tempurl}


\bibitem[Flanagan and Saxe(2001)]%
        {Flanagan2001AvoidingExpExplosionVC}
\bibfield{author}{\bibinfo{person}{Cormac Flanagan} {and}
  \bibinfo{person}{James~B. Saxe}.} \bibinfo{year}{2001}\natexlab{}.
\newblock \showarticletitle{Avoiding exponential explosion: generating compact
  verification conditions}. In \bibinfo{booktitle}{\emph{ACM-SIGACT Symposium
  on Principles of Programming Languages}}.
\newblock


\bibitem[Heljanko et~al\mbox{.}(2005)]%
        {Heljanko2005IncrementalAC}
\bibfield{author}{\bibinfo{person}{Keijo Heljanko}, \bibinfo{person}{Tommi~A.
  Junttila}, {and} \bibinfo{person}{Timo Latvala}.}
  \bibinfo{year}{2005}\natexlab{}.
\newblock \showarticletitle{Incremental and Complete Bounded Model Checking for
  Full PLTL}. In \bibinfo{booktitle}{\emph{International Conference on Computer
  Aided Verification}}.
\newblock


\bibitem[Hoare(1968)]%
        {Hoare1968HoareLogic}
\bibfield{author}{\bibinfo{person}{C.~A.~R. Hoare}.}
  \bibinfo{year}{1968}\natexlab{}.
\newblock \showarticletitle{An Axiomatic Basis for Computer Programming}.
\newblock \bibinfo{journal}{\emph{Commun. ACM}}  \bibinfo{volume}{12}
  (\bibinfo{year}{1968}), \bibinfo{pages}{576--580}.
\newblock
\urldef\tempurl%
\url{https://doi.org/10.1145/363235.363259}
\showDOI{\tempurl}


\bibitem[Jhala and McMillan(2007)]%
        {JhalaM2007ArrayAbstraction}
\bibfield{author}{\bibinfo{person}{Ranjit Jhala} {and}
  \bibinfo{person}{Kenneth~L. McMillan}.} \bibinfo{year}{2007}\natexlab{}.
\newblock \showarticletitle{Array Abstractions from Proofs}. In
  \bibinfo{booktitle}{\emph{Computer Aided Verification, 19th International
  Conference, {CAV} 2007, Berlin, Germany, July 3-7, 2007, Proceedings}}
  \emph{(\bibinfo{series}{Lecture Notes in Computer Science},
  Vol.~\bibinfo{volume}{4590})}, \bibfield{editor}{\bibinfo{person}{Werner
  Damm} {and} \bibinfo{person}{Holger Hermanns}} (Eds.).
  \bibinfo{publisher}{Springer}, \bibinfo{pages}{193--206}.
\newblock
\urldef\tempurl%
\url{https://doi.org/10.1007/978-3-540-73368-3\_23}
\showDOI{\tempurl}


\bibitem[Jung et~al\mbox{.}(2018)]%
        {Jung2018IrisGroundUp}
\bibfield{author}{\bibinfo{person}{Ralf Jung}, \bibinfo{person}{Robbert
  Krebbers}, \bibinfo{person}{Jacques-Henri Jourdan}, \bibinfo{person}{Ales
  Bizjak}, \bibinfo{person}{Lars Birkedal}, {and} \bibinfo{person}{Derek
  Dreyer}.} \bibinfo{year}{2018}\natexlab{}.
\newblock \showarticletitle{Iris from the ground up: A modular foundation for
  higher-order concurrent separation logic}.
\newblock \bibinfo{journal}{\emph{J. Funct. Program.}}  \bibinfo{volume}{28}
  (\bibinfo{year}{2018}), \bibinfo{pages}{e20}.
\newblock
\urldef\tempurl%
\url{https://doi.org/10.1017/S0956796818000151}
\showDOI{\tempurl}


\bibitem[Kroening et~al\mbox{.}(2011)]%
        {DBLP:conf/cav/KroeningOSWW11}
\bibfield{author}{\bibinfo{person}{Daniel Kroening},
  \bibinfo{person}{Jo{\"{e}}l Ouaknine}, \bibinfo{person}{Ofer Strichman},
  \bibinfo{person}{Thomas Wahl}, {and} \bibinfo{person}{James Worrell}.}
  \bibinfo{year}{2011}\natexlab{}.
\newblock \showarticletitle{Linear Completeness Thresholds for Bounded Model
  Checking}. In \bibinfo{booktitle}{\emph{Computer Aided Verification - 23rd
  International Conference, {CAV} 2011, Snowbird, UT, USA, July 14-20, 2011.
  Proceedings}} \emph{(\bibinfo{series}{Lecture Notes in Computer Science},
  Vol.~\bibinfo{volume}{6806})}, \bibfield{editor}{\bibinfo{person}{Ganesh
  Gopalakrishnan} {and} \bibinfo{person}{Shaz Qadeer}} (Eds.).
  \bibinfo{publisher}{Springer}, \bibinfo{pages}{557--572}.
\newblock
\urldef\tempurl%
\url{https://doi.org/10.1007/978-3-642-22110-1\_44}
\showDOI{\tempurl}


\bibitem[Kroening and Strichman(2003)]%
        {Kroening2003EfficientCO}
\bibfield{author}{\bibinfo{person}{Daniel Kroening} {and} \bibinfo{person}{Ofer
  Strichman}.} \bibinfo{year}{2003}\natexlab{}.
\newblock \showarticletitle{Efficient Computation of Recurrence Diameters}. In
  \bibinfo{booktitle}{\emph{International Conference on Verification, Model
  Checking and Abstract Interpretation}}.
\newblock


\bibitem[Mastroeni and Zanardini(2008)]%
        {MastroeniDataDependencyProgramSlicing}
\bibfield{author}{\bibinfo{person}{Isabella Mastroeni} {and}
  \bibinfo{person}{Damiano Zanardini}.} \bibinfo{year}{2008}\natexlab{}.
\newblock \showarticletitle{Data dependencies and program slicing: from syntax
  to abstract semantics}. In \bibinfo{booktitle}{\emph{Proceedings of the 2008
  {ACM} {SIGPLAN} Symposium on Partial Evaluation and Semantics-based Program
  Manipulation, {PEPM} 2008, San Francisco, California, USA, January 7-8,
  2008}}, \bibfield{editor}{\bibinfo{person}{Robert Gl{\"{u}}ck} {and}
  \bibinfo{person}{Oege de~Moor}} (Eds.). \bibinfo{publisher}{{ACM}},
  \bibinfo{pages}{125--134}.
\newblock
\urldef\tempurl%
\url{https://doi.org/10.1145/1328408.1328428}
\showDOI{\tempurl}


\bibitem[Mathur et~al\mbox{.}(2020)]%
        {DBLP:journals/pacmpl/MathurMKMV20}
\bibfield{author}{\bibinfo{person}{Umang Mathur}, \bibinfo{person}{Adithya
  Murali}, \bibinfo{person}{Paul Krogmeier}, \bibinfo{person}{P. Madhusudan},
  {and} \bibinfo{person}{Mahesh Viswanathan}.} \bibinfo{year}{2020}\natexlab{}.
\newblock \showarticletitle{Deciding memory safety for single-pass
  heap-manipulating programs}.
\newblock \bibinfo{journal}{\emph{Proc. {ACM} Program. Lang.}}
  \bibinfo{volume}{4}, \bibinfo{number}{{POPL}} (\bibinfo{year}{2020}),
  \bibinfo{pages}{35:1--35:29}.
\newblock
\urldef\tempurl%
\url{https://doi.org/10.1145/3371103}
\showDOI{\tempurl}


\bibitem[McMillan(2003)]%
        {McMillan2003InterpolationAS}
\bibfield{author}{\bibinfo{person}{Kenneth~L. McMillan}.}
  \bibinfo{year}{2003}\natexlab{}.
\newblock \showarticletitle{Interpolation and SAT-Based Model Checking}. In
  \bibinfo{booktitle}{\emph{International Conference on Computer Aided
  Verification}}.
\newblock


\bibitem[Moser(1990)]%
        {Moser90DataDependencyGraphs}
\bibfield{author}{\bibinfo{person}{Louise~E. Moser}.}
  \bibinfo{year}{1990}\natexlab{}.
\newblock \showarticletitle{Data Dependency Graphs for Ada Programs}.
\newblock \bibinfo{journal}{\emph{{IEEE} Trans. Software Eng.}}
  \bibinfo{volume}{16}, \bibinfo{number}{5} (\bibinfo{year}{1990}),
  \bibinfo{pages}{498--509}.
\newblock
\urldef\tempurl%
\url{https://doi.org/10.1109/32.52773}
\showDOI{\tempurl}


\bibitem[O'Hearn(2019)]%
        {DBLP:journals/cacm/OHearn19}
\bibfield{author}{\bibinfo{person}{Peter~W. O'Hearn}.}
  \bibinfo{year}{2019}\natexlab{}.
\newblock \showarticletitle{Separation logic}.
\newblock \bibinfo{journal}{\emph{Commun. {ACM}}} \bibinfo{volume}{62},
  \bibinfo{number}{2} (\bibinfo{year}{2019}), \bibinfo{pages}{86--95}.
\newblock
\urldef\tempurl%
\url{https://doi.org/10.1145/3211968}
\showDOI{\tempurl}


\bibitem[O'Hearn et~al\mbox{.}(2001)]%
        {OHearn2001LocalRA}
\bibfield{author}{\bibinfo{person}{Peter~W. O'Hearn}, \bibinfo{person}{John~C.
  Reynolds}, {and} \bibinfo{person}{Hongseok Yang}.}
  \bibinfo{year}{2001}\natexlab{}.
\newblock \showarticletitle{Local Reasoning about Programs that Alter Data
  Structures}. In \bibinfo{booktitle}{\emph{CSL}}.
\newblock
\urldef\tempurl%
\url{https://doi.org/10.1007/3-540-44802-0\_1}
\showDOI{\tempurl}


\bibitem[Park et~al\mbox{.}(2000)]%
        {Park2000JavaMC}
\bibfield{author}{\bibinfo{person}{D.Y.W. Park}, \bibinfo{person}{U. Stern},
  \bibinfo{person}{J.U. Skakkebaek}, {and} \bibinfo{person}{D.L. Dill}.}
  \bibinfo{year}{2000}\natexlab{}.
\newblock \showarticletitle{Java model checking}. In
  \bibinfo{booktitle}{\emph{Proceedings ASE 2000. Fifteenth IEEE International
  Conference on Automated Software Engineering}}. \bibinfo{pages}{253--256}.
\newblock
\urldef\tempurl%
\url{https://doi.org/10.1109/ASE.2000.873671}
\showDOI{\tempurl}


\bibitem[Parthasarathy et~al\mbox{.}(2021)]%
        {Parthasarathy2021VCGenerator}
\bibfield{author}{\bibinfo{person}{Gaurav Parthasarathy},
  \bibinfo{person}{Peter M{\"u}ller}, {and} \bibinfo{person}{Alexander~J.
  Summers}.} \bibinfo{year}{2021}\natexlab{}.
\newblock \showarticletitle{Formally Validating a Practical Verification
  Condition Generator}. In \bibinfo{booktitle}{\emph{International Conference
  on Computer Aided Verification}}.
\newblock


\bibitem[Reinhard(2023)]%
        {Reinhard2023ct4ms-TR}
\bibfield{author}{\bibinfo{person}{Tobias Reinhard}.}
  \bibinfo{year}{2023}\natexlab{}.
\newblock \bibinfo{title}{Completeness Thresholds for Memory Safety of Array
  Traversing Programs: Early Technical Report}.
\newblock
\newblock
\showeprint[arxiv]{2211.11885}~[cs.LO]


\bibitem[Reynolds(2002)]%
        {Reynolds2002SeparationLA}
\bibfield{author}{\bibinfo{person}{John~C. Reynolds}.}
  \bibinfo{year}{2002}\natexlab{}.
\newblock \showarticletitle{Separation logic: a logic for shared mutable data
  structures}.
\newblock \bibinfo{journal}{\emph{Proceedings 17th Annual IEEE Symposium on
  Logic in Computer Science}} (\bibinfo{year}{2002}), \bibinfo{pages}{55--74}.
\newblock
\urldef\tempurl%
\url{https://doi.org/10.1109/LICS.2002.1029817}
\showDOI{\tempurl}


\bibitem[Sheeran et~al\mbox{.}(2000)]%
        {DBLP:conf/fmcad/SheeranSS00}
\bibfield{author}{\bibinfo{person}{Mary Sheeran}, \bibinfo{person}{Satnam
  Singh}, {and} \bibinfo{person}{Gunnar St{\aa}lmarck}.}
  \bibinfo{year}{2000}\natexlab{}.
\newblock \showarticletitle{Checking Safety Properties Using Induction and a
  SAT-Solver}. In \bibinfo{booktitle}{\emph{Formal Methods in Computer-Aided
  Design, Third International Conference, {FMCAD} 2000, Austin, Texas, USA,
  November 1-3, 2000, Proceedings}} \emph{(\bibinfo{series}{Lecture Notes in
  Computer Science}, Vol.~\bibinfo{volume}{1954})},
  \bibfield{editor}{\bibinfo{person}{Warren A.~Hunt Jr.} {and}
  \bibinfo{person}{Steven~D. Johnson}} (Eds.). \bibinfo{publisher}{Springer},
  \bibinfo{pages}{108--125}.
\newblock
\urldef\tempurl%
\url{https://doi.org/10.1007/3-540-40922-X\_8}
\showDOI{\tempurl}


\bibitem[Weiser(1984)]%
        {Weiser84ProgramSlicing}
\bibfield{author}{\bibinfo{person}{Mark~D. Weiser}.}
  \bibinfo{year}{1984}\natexlab{}.
\newblock \showarticletitle{Program Slicing}.
\newblock \bibinfo{journal}{\emph{{IEEE} Trans. Software Eng.}}
  \bibinfo{volume}{10}, \bibinfo{number}{4} (\bibinfo{year}{1984}),
  \bibinfo{pages}{352--357}.
\newblock
\urldef\tempurl%
\url{https://doi.org/10.1109/TSE.1984.5010248}
\showDOI{\tempurl}


\end{thebibliography}
	
\end{document}